\expandafter \def \csname CHAPLABELintro\endcsname {1}
\expandafter \def \csname EQLABELduality\endcsname {1.1?}
\expandafter \def \csname CHAPLABELToric\endcsname {2}
\expandafter \def \csname FIGLABELrays\endcsname {2.1?}
\expandafter \def \csname CHAPLABELmatter\endcsname {3}
\expandafter \def \csname TABLABELcharged\endcsname {3.1?}
\expandafter \def \csname EQLABELanomaly\endcsname {3.1?}
\expandafter \def \csname EQLABELtotalmatter\endcsname {3.2?}
\expandafter \def \csname EQLABELmultmatter\endcsname {3.3?}
\expandafter \def \csname FIGLABELfan\endcsname {3.1?}
\expandafter \def \csname CHAPLABELeg\endcsname {4}
\expandafter \def \csname CHAPLABELfin\endcsname {5}
%%%%%%%%%%%%%%%%%%%%%%%%%%%%%%%%%%%%%%%%%%%%%%%%%%%%%%%%%%%%%%%%%%%%%%%%%%%%%%%
%Fonts%%%%%%%%%%%%%%%%%%%%%%%%%%%%%%%%%%%%%%%%%%%%%%%%%%%%%%%%%%%%%%%%%%%%%%%%%
%%%%%%%%%%%%%%%%%%%%%%%%%%%%%%%%%%%%%%%%%%%%%%%%%%%%%%%%%%%%%%%%%%%%%%%%%%%%%%%

\font\eightrm=cmr8 at 8pt

\font\seventeenrm=cmr17 at 17pt
\font\twentyonerm=cmr17 at 21pt

\font\ss=cmss10

\font\csc=cmcsc10

\font\twelvecal=cmsy10 at 12pt

\font\twelvemath=cmmi12

\font\seventeenbold=cmbx7 at 17pt

\font\fively=lasy5
\font\sevenly=lasy7
\font\tenly=lasy10

\textfont10=\tenly
\scriptfont10=\sevenly
\scriptscriptfont10=\fively
%%%%%%%%%%%%%%%%%%%%%%%%%%%%%%%%%%%%%%%%%%%%%%%%%%%%%%%%%%%%%%%%%%%%%%%%%%%%%%%
%Formatting%%%%%%%%%%%%%%%%%%%%%%%%%%%%%%%%%%%%%%%%%%%%%%%%%%%%%%%%%%%%%%%%%%%%
%%%%%%%%%%%%%%%%%%%%%%%%%%%%%%%%%%%%%%%%%%%%%%%%%%%%%%%%%%%%%%%%%%%%%%%%%%%%%%%
\magnification=1200
\parskip=10pt
\parindent=20pt
\def\today{\ifcase\month\or January\or February\or March\or April\or May\or
June
       \or July\or August\or September\or October\or November\or December\fi
       \space\number\day, \number\year}

\def\title#1{\footline={\ifnum\pageno<2\hfil
       \else\hss\tenrm\folio\hss\fi}\vskip1truein\centerline{{#1}
       \footnote{\raise1ex\hbox{*}}{\eightrm Supported in part
       by the Robert A. Welch Foundation and N.S.F. Grants
       PHY-880637 and\break PHY-8605978.}}}

\def\newpage{\vfill\eject}
\def\abstract#1{\centerline{\bf ABSTRACT}\vskip.2truein{\narrower\noindent#1
       \smallskip}}

\def\runninghead#1#2{\voffset=2\baselineskip\nopagenumbers
       \headline={\ifodd\pageno\rightheadline\else \leftheadline\fi}
       \def\rightheadline{{\sl#1}\hfill{\rm\folio}}
       \def\leftheadline{{\rm\folio}\hfill{\sl#2}}}
\def\SS{\mathhexbox278}

\newcount\footnoteno
\def\Footnote#1{\advance\footnoteno by 1
                \let\SF=\empty
                \ifhmode\edef\SF{\spacefactor=\the\spacefactor}\/\fi
                $^{\the\footnoteno}$\ignorespaces
                \SF\vfootnote{$^{\the\footnoteno}$}{#1}}

\def\figbox#1#2#3{\vbox{\vskip15pt
                   \vbox{\hrule
                    \hbox{\vrule
                     \vbox{\vskip12truept\centerline #1 \vskip6truept
                          {\hskip.4truein\vbox{\hsize=5truein\noindent
                          {\bf Figure\hskip5truept#2:}\hskip5truept#3}}
                     \vskip18truept}
                    \vrule}
                   \hrule}}}
\def\place#1#2#3{\vbox to0pt{\kern-\parskip\kern-7pt
                             \kern-#2truein\hbox{\kern#1truein #3}
                             \vss}\nointerlineskip}
\def\figurecaption#1#2{\kern.75truein\vbox{\hsize=5truein\noindent{\bf Figure
    \figlabel{#1}:} #2}}
\def\tablecaption#1#2{\kern.75truein\lower12truept\hbox{\vbox{\hsize=5truein
    \noindent{\bf Table\hskip5truept\tablabel{#1}:} #2}}}
\def\boxed#1{\lower3pt\hbox{
                       \vbox{\hrule\hbox{\vrule

\vbox{\kern2pt\hbox{\kern3pt#1\kern3pt}\kern3pt}\vrule}
                         \hrule}}}
%%%%%%%%%%%%%%%%%%%%%%%%%%%%%%%%%%%%%%%%%%%%%%%%%%%%%%%%%%%%%%%%%%%%%%%%%%%%%%%
%Greek characters%%%%%%%%%%%%%%%%%%%%%%%%%%%%%%%%%%%%%%%%%%%%%%%%%%%%%%%%%%%%%%
%%%%%%%%%%%%%%%%%%%%%%%%%%%%%%%%%%%%%%%%%%%%%%%%%%%%%%%%%%%%%%%%%%%%%%%%%%%%%%%

\def\g{\gamma}
\def\D{\Delta}

\def\l{\lambda}\def\L{\Lambda}

\def\P{\Pi}

%%%%%%%%%%%%%%%%%%%%%%%%%%%%%%%%%%%%%%%%%%%%%%%%%%%%%%%%%%%%%%%%%%%%%%%%%%%%%%%
%Calligraphic capitals%%%%%%%%%%%%%%%%%%%%%%%%%%%%%%%%%%%%%%%%%%%%%%%%%%%%%%%%%
%%%%%%%%%%%%%%%%%%%%%%%%%%%%%%%%%%%%%%%%%%%%%%%%%%%%%%%%%%%%%%%%%%%%%%%%%%%%%%%
\def\ca#1{\relax\ifmmode {{\cal #1}}\else $\cal #1$\fi}

\def\calb{{\cal B}}

\def\calm{{\cal M}}

%%%%%%%%%%%%%%%%%%%%%%%%%%%%%%%%%%%%%%%%%%%%%%%%%%%%%%%%%%%%%%%%%%%%%%%%%%%%%%%
% Poor man's Blackboard Bold%%%%%%%%%%%%%%%%%%%%%%%%%%%%%%%%%%%%%%%%%%%%%%%%%%%
%%%%%%%%%%%%%%%%%%%%%%%%%%%%%%%%%%%%%%%%%%%%%%%%%%%%%%%%%%%%%%%%%%%%%%%%%%%%%%%
\def\inbar{\vrule height1.5ex width.4pt depth0pt}
\def\IB{\relax{\rm I\kern-.18em B}}
\def\IC{\relax\hbox{\kern.25em$\inbar\kern-.3em{\rm C}$}}
\def\ID{\relax{\rm I\kern-.18em D}}
\def\IE{\relax{\rm I\kern-.18em E}}
\def\IF{\relax{\rm I\kern-.18em F}}
\def\IG{\relax\hbox{\kern.25em$\inbar\kern-.3em{\rm G}$}}
\def\IH{\relax{\rm I\kern-.18em H}}
\def\II{\relax{\rm I\kern-.18em I}}
\def\IK{\relax{\rm I\kern-.18em K}}
\def\IL{\relax{\rm I\kern-.18em L}}
\def\IM{\relax{\rm I\kern-.18em M}}
\def\IN{\relax{\rm I\kern-.18em N}}
\def\IO{\relax\hbox{\kern.25em$\inbar\kern-.3em{\rm O}$}}
\def\IP{\relax{\rm I\kern-.18em P}}
\def\IQ{\relax\hbox{\kern.25em$\inbar\kern-.3em{\rm Q}$}}
\def\IR{\relax{\rm I\kern-.18em R}}
\def\IZ{\relax\ifmmode\hbox{\ss Z\kern-.4em Z}\else{\ss Z\kern-.4em Z}\fi}
\def\IGa{\relax{\rm I}\kern-.18em\Gamma}
\def\IPi{\relax{\rm I}\kern-.18em\Pi}
\def\ITh{\relax\hbox{\kern.25em$\inbar\kern-.3em\Theta$}}
\def\IOm{\relax\thinspace\inbar\kern1.95pt\inbar\kern-5.525pt\Omega}

%Papers, Lecture Notes on Complex Manifolds etc.

\def\ie{{\it i.e.,\ \/}}

\def\cy{Calabi--Yau}
\def\cym{Calabi--Yau manifold}
\def\cys{Calabi--Yau manifolds}
\def\cyt{Calabi--Yau threefold}

\def\H#1#2{\relax\ifmmode {H^{#1#2}}\else $H^{#1 #2}$\fi}
\def\M{\relax\ifmmode{\calm}\else $\calm$\fi}

\def\Bigcheck{\lower3.8pt\hbox{\smash{\hbox{{\twentyonerm \v{}}}}}}
\def\bigboldcheck{\smash{\hbox{{\seventeenbold\v{}}}}}

\def\Bighat{\lower3.8pt\hbox{\smash{\hbox{{\twentyonerm \^{}}}}}}

\def\Msharp{\relax\ifmmode{\calm^\sharp}\else $\smash{\calm^\sharp}$\fi}
\def\Mflat{\relax\ifmmode{\calm^\flat}\else $\smash{\calm^\flat}$\fi}
\def\preMcheck{\kern2pt\hbox{\Bigcheck\kern-12pt{$\cal M$}}}
\def\Mcheck{\relax\ifmmode\preMcheck\else $\preMcheck$\fi}
\def\preMhat{\kern2pt\hbox{\Bighat\kern-12pt{$\cal M$}}}
\def\Mhat{\relax\ifmmode\preMhat\else $\preMhat$\fi}

\def\Bsharp{\relax\ifmmode{\calb^\sharp}\else $\calb^\sharp$\fi}
\def\Bflat{\relax\ifmmode{\calb^\flat}\else $\calb^\flat$ \fi}
\def\preBcheck{\hbox{\Bigcheck\kern-9pt{$\cal B$}}}
\def\Bcheck{\relax\ifmmode\preBcheck\else $\preBcheck$\fi}
\def\preBhat{\hbox{\Bighat\kern-9pt{$\cal B$}}}
\def\Bhat{\relax\ifmmode\preBhat\else $\preBhat$\fi}

\def\figBcheck{\kern3pt\hbox{\raise1pt\hbox{\bigboldcheck}\kern-11pt
    {\twelvecal B}}}
\def\figBsharp{{\twelvecal B}\raise5pt\hbox{$\twelvemath\sharp$}}
\def\figBflat{{\twelvecal B}\raise5pt\hbox{$\twelvemath\flat$}}

\def\gcheck{\hbox{\lower2.5pt\hbox{\Bigcheck}\kern-8pt$\g$}}
\def\lhat{\hbox{\raise.5pt\hbox{\Bighat}\kern-8pt$\l$}}

\def\Fcheck{\kern2pt\hbox{\raise1pt\hbox{\Bigcheck}\kern-10pt{$\cal F$}}}
\def\Fhat{\kern2pt\hbox{\raise1pt\hbox{\Bighat}\kern-10pt{$\cal F$}}}

\def\cp#1{\relax\ifmmode {\IP\kern-2pt{}_{#1}}\else $\IP\kern-2pt{}_{#1}$\fi}
\def\h#1#2{\relax\ifmmode {b_{#1#2}}\else $b_{#1#2}$\fi}

\def\frac#1#2{{#1\over #2}}

\def\cone{\relax\thinspace\hbox{$<\kern-.8em{)}$}}
\mathchardef\mho"0A30

\def\-{\hphantom{-}}

%References

\def\npb#1{Nucl.\ Phys.\ {\bf B#1}}

\def\plb#1{Phys. Lett. {\bf #1B}}

% Pictures

\def\picture #1 by #2 (#3){\vbox to #2{\hrule width #1 height 0pt depth 0pt
                                       \vfill\special{picture #3}}}
\def\scaledpicture #1 by #2 (#3 scaled #4){{\dimen0=#1 \dimen1=#2
           \divide\dimen0 by 1000 \multiply\dimen0 by #4
            \divide\dimen1 by 1000 \multiply\dimen1 by #4
            \picture \dimen0 by \dimen1 (#3 scaled #4)}}
\def\illustration #1 by #2 (#3){\vbox to #2{\hrule width #1 height 0pt depth
0pt
                                       \vfill\special{illustration #3}}}
\def\scaledillustration #1 by #2 (#3 scaled #4){{\dimen0=#1 \dimen1=#2
           \divide\dimen0 by 1000 \multiply\dimen0 by #4
            \divide\dimen1 by 1000 \multiply\dimen1 by #4
            \illustration \dimen0 by \dimen1 (#3 scaled #4)}}

%Letters, Letters of Recommendation, Referee's Reports, Itineraries, etc.

\def\delaOssa{\nobreak\vskip1truein\hbox to\hsize
       {\hskip 4truein Xenia de la Ossa\hfill}}

\def\hoy{\number\day\space de \ifcase\month\or enero\or febrero\or marzo\or
       abril\or mayo\or junio\or julio\or agosto\or septiembre\or octubre\or
       noviembre\or diciembre\fi\space de \number\year}

%%%%%%%%%%%%%%%%%%%%%%%%%%%%%%%%%%%%%%%%%%%%%%%%%%%%%%%%%%%%%%%%%%%%%%%%%%%%%%%
%Equation macros%%%%%%%%%%%%%%%%%%%%%%%%%%%%%%%%%%%%%%%%%%%%%%%%%%%%%%%%%%%%%%%
%%%%%%%%%%%%%%%%%%%%%%%%%%%%%%%%%%%%%%%%%%%%%%%%%%%%%%%%%%%%%%%%%%%%%%%%%%%%%%%
%For use with \eqalign

\newif\ifproofmode
\proofmodefalse

\newif\ifforwardreference
\forwardreferencefalse

\newif\ifchapternumbers
\chapternumbersfalse

\newif\ifcontinuousnumbering
\continuousnumberingfalse

\newif\iffigurechapternumbers
\figurechapternumbersfalse

\newif\ifcontinuousfigurenumbering
\continuousfigurenumberingfalse

\newif\iftablechapternumbers
\tablechapternumbersfalse

\newif\ifcontinuoustablenumbering
\continuoustablenumberingfalse

\font\eqsixrm=cmr6

\def\marginstyle{\eqsixrm}

\newtoks\chapletter
\newcount\chapno
\newcount\eqlabelno
\newcount\figureno
\newcount\tableno

\chapno=0
\eqlabelno=0
\figureno=0
\tableno=0

\def\chapfolio{\ifnum\chapno>0 \the\chapno\else\the\chapletter\fi}

\def\bumpchapno{\ifnum\chapno>-1 \global\advance\chapno by 1
\else\global\advance\chapno by -1 \setletter\chapno\fi
\ifcontinuousnumbering\else\global\eqlabelno=0 \fi
\ifcontinuousfigurenumbering\else\global\figureno=0 \fi
\ifcontinuoustablenumbering\else\global\tableno=0 \fi}

\def\setletter#1{\ifcase-#1{}\or{}%
\or\global\chapletter={A}%
\or\global\chapletter={B}%
\or\global\chapletter={C}%
\or\global\chapletter={D}%
\or\global\chapletter={E}%
\or\global\chapletter={F}%
\or\global\chapletter={G}%
\or\global\chapletter={H}%
\or\global\chapletter={I}%
\or\global\chapletter={J}%
\or\global\chapletter={K}%
\or\global\chapletter={L}%
\or\global\chapletter={M}%
\or\global\chapletter={N}%
\or\global\chapletter={O}%
\or\global\chapletter={P}%
\or\global\chapletter={Q}%
\or\global\chapletter={R}%
\or\global\chapletter={S}%
\or\global\chapletter={T}%
\or\global\chapletter={U}%
\or\global\chapletter={V}%
\or\global\chapletter={W}%
\or\global\chapletter={X}%
\or\global\chapletter={Y}%
\or\global\chapletter={Z}\fi}

\def\tempsetletter#1{\ifcase-#1{}\or{}%
\or\global\chapletter={A}%
\or\global\chapletter={B}%
\or\global\chapletter={C}%
\or\global\chapletter={D}%
\or\global\chapletter={E}%
\or\global\chapletter={F}%
\or\global\chapletter={G}%
\or\global\chapletter={H}%
\or\global\chapletter={I}%
\or\global\chapletter={J}%
\or\global\chapletter={K}%
\or\global\chapletter={L}%
\or\global\chapletter={M}%
\or\global\chapletter={N}%
\or\global\chapletter={O}%
\or\global\chapletter={P}%
\or\global\chapletter={Q}%
\or\global\chapletter={R}%
\or\global\chapletter={S}%
\or\global\chapletter={T}%
\or\global\chapletter={U}%
\or\global\chapletter={V}%
\or\global\chapletter={W}%
\or\global\chapletter={X}%
\or\global\chapletter={Y}%
\or\global\chapletter={Z}\fi}

\def\chapshow#1{\ifnum#1>0 \relax#1%
\else{\tempsetletter{\number#1}\chapno=#1\chapfolio}\fi}

\def\chaplabel#1{\bumpchapno\ifproofmode\ifforwardreference
\immediate\write1{\noexpand\expandafter\noexpand\def
\noexpand\csname CHAPLABEL#1\endcsname{\the\chapno}}\fi\fi
\global\expandafter\edef\csname CHAPLABEL#1\endcsname
{\the\chapno}\ifproofmode\llap{\hbox{\marginstyle #1\ }}\fi\chapfolio}

\def\eqnum{\global\advance\eqlabelno by 1
\eqno(\ifchapternumbers\chapfolio.\fi\the\eqlabelno)}

\def\eqlabel#1{\global\advance\eqlabelno by 1 \ifproofmode\ifforwardreference
\immediate\write1{\noexpand\expandafter\noexpand\def
\noexpand\csname EQLABEL#1\endcsname{\the\chapno.\the\eqlabelno?}}\fi\fi
\global\expandafter\edef\csname EQLABEL#1\endcsname
{\the\chapno.\the\eqlabelno?}\eqno(\ifchapternumbers\chapfolio.\fi
\the\eqlabelno)\ifproofmode\rlap{\hbox{\marginstyle #1}}\fi}

\def\eqalignnum{\global\advance\eqlabelno by 1
&(\ifchapternumbers\chapfolio.\fi\the\eqlabelno)}

\def\eqalignlabel#1{\global\advance\eqlabelno by 1 \ifproofmode
\ifforwardreference\immediate\write1{\noexpand\expandafter\noexpand\def
\noexpand\csname EQLABEL#1\endcsname{\the\chapno.\the\eqlabelno?}}\fi\fi
\global\expandafter\edef\csname EQLABEL#1\endcsname
{\the\chapno.\the\eqlabelno?}&(\ifchapternumbers\chapfolio.\fi
\the\eqlabelno)\ifproofmode\rlap{\hbox{\marginstyle #1}}\fi}

\def\eqref#1{\hbox{(\ifundefined{EQLABEL#1}***)\ifproofmode\ifforwardreference%
\else\write16{ ***Undefined Equation Reference #1*** }\fi
\else\write16{ ***Undefined Equation Reference #1*** }\fi
\else\edef\LABxx{\getlabel{EQLABEL#1}}%
\def\LAByy{\expandafter\stripchap\LABxx}\ifchapternumbers%
\chapshow{\LAByy}.\expandafter\stripeq\LABxx%
\else\ifnum\number\LAByy=\chapno\relax\expandafter\stripeq\LABxx%
\else\chapshow{\LAByy}.\expandafter\stripeq\LABxx\fi\fi)\fi}%
\ifproofmode\write2{Equation #1}\fi}

\def\fignum{\global\advance\figureno by 1
\relax\iffigurechapternumbers\chapfolio.\fi\the\figureno}

\def\figlabel#1{\global\advance\figureno by 1
\relax\ifproofmode\ifforwardreference
\immediate\write1{\noexpand\expandafter\noexpand\def
\noexpand\csname FIGLABEL#1\endcsname{\the\chapno.\the\figureno?}}\fi\fi
\global\expandafter\edef\csname FIGLABEL#1\endcsname
{\the\chapno.\the\figureno?}\iffigurechapternumbers\chapfolio.\fi
\ifproofmode\llap{\hbox{\marginstyle#1
\kern1.2truein}}\relax\fi\the\figureno}

% THE FOLLOWING LINE CANNOT BE BROKEN BEFORE 80 CHAR
\def\figref#1{\hbox{\ifundefined{FIGLABEL#1}!!!!\ifproofmode\ifforwardreference%
\else\write16{ ***Undefined Figure Reference #1*** }\fi
\else\write16{ ***Undefined Figure Reference #1*** }\fi
\else\edef\LABxx{\getlabel{FIGLABEL#1}}%
\def\LAByy{\expandafter\stripchap\LABxx}\iffigurechapternumbers%
\chapshow{\LAByy}.\expandafter\stripeq\LABxx%
\else\ifnum \number\LAByy=\chapno\relax\expandafter\stripeq\LABxx%
\else\chapshow{\LAByy}.\expandafter\stripeq\LABxx\fi\fi\fi}%
\ifproofmode\write2{Figure #1}\fi}

\def\tabnum{\global\advance\tableno by 1
\relax\iftablechapternumbers\chapfolio.\fi\the\tableno}

\def\tablabel#1{\global\advance\tableno by 1
\relax\ifproofmode\ifforwardreference
\immediate\write1{\noexpand\expandafter\noexpand\def
\noexpand\csname TABLABEL#1\endcsname{\the\chapno.\the\tableno?}}\fi\fi
\global\expandafter\edef\csname TABLABEL#1\endcsname
{\the\chapno.\the\tableno?}\iftablechapternumbers\chapfolio.\fi
\ifproofmode\llap{\hbox{\marginstyle#1
\kern1.2truein}}\relax\fi\the\tableno}

% THE FOLLOWING LINE CANNOT BE BROKEN BEFORE 80 CHAR
\def\tabref#1{\hbox{\ifundefined{TABLABEL#1}!!!!\ifproofmode\ifforwardreference%
\else\write16{ ***Undefined Table Reference #1*** }\fi
\else\write16{ ***Undefined Table Reference #1*** }\fi
\else\edef\LABtt{\getlabel{TABLABEL#1}}%
\def\LABTT{\expandafter\stripchap\LABtt}\iftablechapternumbers%
\chapshow{\LABTT}.\expandafter\stripeq\LABtt%
\else\ifnum\number\LABTT=\chapno\relax\expandafter\stripeq\LABtt%
\else\chapshow{\LABTT}.\expandafter\stripeq\LABtt\fi\fi\fi}%
\ifproofmode\write2{Table#1}\fi}

\newdimen\sectionskip     \sectionskip=20truept
\newcount\sectno
\def\section#1#2{\sectno=0 \null\vskip\sectionskip
    \centerline{\chaplabel{#1}.~~{\bf#2}}\nobreak\vskip.2truein
    \noindent\ignorespaces}

\def\advancesectno{\global\advance\sectno by 1}
\def\sectfolio{\number\sectno}
\def\subsection#1{\goodbreak\advancesectno\null\vskip10pt
                  \noindent\chapfolio.~\sectfolio.~{\bf #1}
                  \nobreak\vskip.05truein\noindent\ignorespaces}

\def\uttg#1{\null\vskip.1truein
    \ifproofmode \line{\hfill{\bf Draft}:
    UTTG--{#1}--\number\year}\line{\hfill\today}
    \else \line{\hfill UTTG--{#1}--\number\year}
    \line{\hfill\ifcase\month\or January\or February\or March\or April\or
May\or June
    \or July\or August\or September\or October\or November\or December\fi
    \space\number\year}\fi}

\def\getlabel#1{\csname#1\endcsname}
\def\ifundefined#1{\expandafter\ifx\csname#1\endcsname\relax}
\def\stripchap#1.#2?{#1}
\def\stripeq#1.#2?{#2}

%%%%%%%%%%%%%%%%%%%%%%%%%%%%%%%%%%%%%%%%%%%%%%%%%%%%%%%%%%%%%%%%%%%%%%%%%%%%%%
%Reference macros%%%%%%%%%%%%%%%%%%%%%%%%%%%%%%%%%%%%%%%%%%%%%%%%%%%%%%%%%%%%%
%%%%%%%%%%%%%%%%%%%%%%%%%%%%%%%%%%%%%%%%%%%%%%%%%%%%%%%%%%%%%%%%%%%%%%%%%%%%%%
%
\catcode`@=11 % This allows us to modify PLAIN macros.
\def\space@ver#1{\let\@sf=\empty\ifmmode#1\else\ifhmode%
\edef\@sf{\spacefactor=\the\spacefactor}\unskip${}#1$\relax\fi\fi}
\newcount\referencecount     \referencecount=0
\newif\ifreferenceopen       \newwrite\referencewrite
\newtoks\rw@toks
\def\refmark#1{\relax[#1]}
\def\refend{\refmark{\number\referencecount}}
\newcount\lastrefsbegincount \lastrefsbegincount=0
\def\refsend{\refmark{\count255=\referencecount%
\advance\count255 by -\lastrefsbegincount%
\ifcase\count255 \number\referencecount%
\or\number\lastrefsbegincount,\number\referencecount%
\else\number\lastrefsbegincount-\number\referencecount\fi}}
\def\refch@ck{\chardef\rw@write=\referencewrite
\ifreferenceopen\else\referenceopentrue
\immediate\openout\referencewrite=referenc.texauxil \fi}
%
% In \obeyendofline, we say `\let^^M=\relax
{\catcode`\^^M=\active % these lines must end with %
  \gdef\obeyendofline{\catcode`\^^M\active \let^^M\ }}%
%
% In \ignoreendofline, we say `\let^^M=\relax
{\catcode`\^^M=\active % these lines must end with %
  \gdef\ignoreendofline{\catcode`\^^M=5}}
{\obeyendofline\gdef\rw@start#1{\def\t@st{#1}\ifx\t@st\blankend%
\endgroup\@sf\relax\else\ifx\t@st\bl@nkend\endgroup\@sf\relax%
\else\rw@begin#1
\backtotext
\fi\fi}}
{\obeyendofline\gdef\rw@begin#1
{\def\n@xt{#1}\rw@toks={#1}\relax%
\rw@next}}
\def\blankend{}
{\obeylines\gdef\bl@nkend{
}}
\newif\iffirstrefline  \firstreflinetrue
\def\rwr@teswitch{\ifx\n@xt\blankend\let\n@xt=\rw@begin%
\else\iffirstrefline\global\firstreflinefalse%
\immediate\write\rw@write{\noexpand\obeyendofline\the\rw@toks}%
\let\n@xt=\rw@begin%
\else\ifx\n@xt\rw@@d \def\n@xt{\immediate\write\rw@write{%
\noexpand\ignoreendofline}\endgroup\@sf}%
\else\immediate\write\rw@write{\the\rw@toks}%
\let\n@xt=\rw@begin\fi\fi\fi}
\def\rw@next{\rwr@teswitch\n@xt}
\def\rw@@d{\backtotext} \let\rw@end=\relax
\let\backtotext=\relax

\newdimen\refindent     \refindent=30pt
\def\Textindent#1{\noindent\llap{#1\enspace}\ignorespaces}
\def\refitem#1{\par\hangafter=0 \hangindent=\refindent\Textindent{#1}}
\def\REFNUM#1{\space@ver{}\refch@ck\firstreflinetrue%
\global\advance\referencecount by 1 \xdef#1{\the\referencecount}}
\def\refnum#1{\space@ver{}\refch@ck\firstreflinetrue%
\global\advance\referencecount by 1\xdef#1{\the\referencecount}\refend}

\def\REF#1{\REFNUM#1%
\immediate\write\referencewrite{%
\noexpand\refitem{#1.}}%
\begingroup\obeyendofline\rw@start}
\def\ref{\refnum\?%
\immediate\write\referencewrite{\noexpand\refitem{\?.}}%
\begingroup\obeyendofline\rw@start}
\def\Ref#1{\refnum#1%
\immediate\write\referencewrite{\noexpand\refitem{#1.}}%
\begingroup\obeyendofline\rw@start}
\def\REFS#1{\REFNUM#1\global\lastrefsbegincount=\referencecount%
\immediate\write\referencewrite{\noexpand\refitem{#1.}}%
\begingroup\obeyendofline\rw@start}

\def\REFSCON#1{\REF#1}

\def\cite#1{\refmark#1}
\catcode`@=12 % at signs are no longer letters
%
%%%%%%%%%%%%%%%%%%%%%%%%%%%%%%%%%%%%%%%%%%%%%%%%%%%%%%%%%%%%%%%%%%%%%%%%%%%%%%%
%\dump

%\input xenia.mac
\input epsf.tex
%\proofmodetrue
\baselineskip=15pt plus 1pt minus 1pt
\parskip=5pt
\chapternumberstrue
%\forwardreferencetrue
\figurechapternumberstrue
\tablechapternumberstrue
\ifproofmode
\immediate\openout2=allcrossreferfile \fi
\ifforwardreference\input labelfile
\ifproofmode\immediate\openout1=labelfile \fi\fi
%\noblackboxes
\hfuzz=2pt
\vfuzz=3pt

%%%%%%%%%%%%%%%%%%%%%%%%%%%%%%%%%%%%%%%%%%%%%%%%%%%%%%%%%%%%%%%%%%%%%%%%%%%%%%%
%macros

\def\hourandminute{\count255=\time\divide\count255 by 60
\xdef\hour{\number\count255}
\multiply\count255 by -60\advance\count255 by\time
\hour:\ifnum\count255<10 0\fi\the\count255}

\def\subsection#1{\goodbreak\advancesectno\null\vskip10pt
                  \noindent{\it \chapfolio.\sectfolio.~#1}
                  \nobreak\vskip.05truein\noindent\ignorespaces}

\def\subsubsection#1#2{\vskip5pt\goodbreak
                       \noindent\vbox{\hbox{\rm #1~\it #2}\vskip2pt\hrule}
                       \nobreak\vskip.03truein\noindent\ignorespaces}

\def\cite#1{\refmark{#1}}

\def\\{\hfill\break}

%For use with \eqalign

\def\point#1{\noindent\setbox0=\hbox{#1}\kern-\wd0\box0}

\def\cyt{Calabi-Yau threefold}

\def\etal{{\it et al.\/}}

%%%%%%%%%%%%%%%%%%%%%%%%%%%%%%%%%%%%%%%%%%%%%%%%%%%%%%%%%
%%%%%%%[ Title Page ]%%%%%%%%%%%%%%%%%%%%%%%%%%%%%%%%%%%%
%%%%%%%%%%%%%%%%%%%%%%%%%%%%%%%%%%%%%%%%%%%%%%%%%%%%%%%%%
\nopagenumbers\pageno=0
\rightline{\eightrm UTTG-17-97}\vskip-5pt
\rightline{\eightrm hep-th/9707049}\vskip-5pt
\rightline{\eightrm July 3, 1997}

\vskip1.2truein
\centerline{\seventeenrm Matter from Toric Geometry}
\vskip20pt
\centerline{\csc Philip~Candelas$^1$, Eugene~Perevalov$^2$ and 
Govindan~Rajesh$^3$}
\vfootnote{$^{\eightrm 1}$}{\eightrm candelas@physics.utexas.edu}
\vfootnote{$^{\eightrm 2}$}{\eightrm pereval@physics.utexas.edu}
\vfootnote{$^{\eightrm 3}$}{\eightrm rajesh@physics.utexas.edu}

\vskip.8truein\bigskip
\centerline{\it Theory Group}
\centerline{\it Department of Physics}
\centerline{\it University of Texas}
\centerline{\it Austin, TX 78712, USA}
\vskip1.3in\bigskip
\nobreak\vbox{
\centerline{\bf ABSTRACT}
\vskip.4truein
\noindent{We present an algorithm for obtaining the matter content of
effective six-dimensional theories resulting from compactification of F-theory
on elliptic \cyt s which are hypersurfaces in toric varieties. The algorithm
allows us to read off the matter content of the theory from the polyhedron
describing the \cym . This is based on the generalized Green-Schwarz anomaly
cancellation condition.}
} 
\newpage
%%%%%%%%%%%%%%%%%%%%%%%%%[Table of Contents]%%%%%%%%%%%%%%%%%%%%%%%%%%%
    {\bf Contents}
\vskip5pt

    1. Introduction 
\vskip3pt

    2. Some Results in Toric Geometry
\vskip3pt

    3. Identifying the Matter Content
\vskip3pt

    4. Examples
\vskip3pt

\hskip10pt {\it 4.1 The mirrors of models with }{$n=0$}
\vskip3pt
\hskip10pt {\it 4.2 The mirrors of models with higher values of }{$n$} 
\vskip3pt

    5. Discussion
\vskip3pt

\newpage
%%%%%%%%%%%%%%%%%%%%%%%%%%%[ The Article ]%%%%%%%%%%%%%%%%%%%%%%%%%%%%%
\pageno=1
\headline={\ifproofmode\hfil\eightrm draft:\ \today\ \hourandminute\else\hfil\fi}
\footline={\rm\hfil\folio\hfil}
\section{intro}{Introduction}
The dualities of String Theory have been the subject of extensive study during
the last two years. Of particular interest to us here is the duality~ 
\REFS\rCV{S. Kachru and C. Vafa, Nucl. Phys. {\bf B450} (1995) 69, 
hep-th/9505105.}
\REFSCON\rFSHV{S. Ferrara, J. Harvey, A. Strominger and C. Vafa,\\
 Phys. Lett. {\bf 361B} (1995) 59, hep-th/9505162.}
\refsend\
$$\hbox{Het}[K3\times T^2, G] = \hbox{IIA}[\ca{M}] \eqlabel{duality}$$
between a $(0,4)$ heterotic compactification on $K3\times T^2$ with gauge
group $G$, and a type IIA compactification on a \cym, \ca{M}~
\REFS\rKLM{A. Klemm, W. Lerche and P. Mayr, Phys. Lett. {\bf 357B} (1995) 
113, hep-th/9506112.}
\REFSCON\rVW{C. Vafa and E. Witten, hep-th/9507050.}
\REFSCON\rKLT{V. Kaplunovsky, J. Louis and S. Theisen,\\ Phys. Lett. {\bf 357B}
(1995) 71, hep-th/9506110.}
\REFSCON\rKKLMV{S. Kachru, A. Klemm, W. Lerche, P. Mayr and C. Vafa,\\
Nucl. Phys. {\bf B459} (1996) 537, hep-th/9508155.}
\refsend.

In many relevant cases, the \cys\ can be conveniently
represented in terms of the toric data as was shown in~
\REFS\CF{P.~Candelas and A.~Font, hep-th/9603170.}
\REFSCON\BS{P.~Candelas, E.~Perevalov and G.~Rajesh, hep-th/9606133.}
\REFSCON\BC{P.~Candelas, E.~Perevalov and G.~Rajesh, hep-th/9703148.}
\refsend
(see also the important articles
\REFS\MVI{D.~R.~Morrison and C.~Vafa, Nucl. Phys. {\bf B473} (1996) 74,
hep-th/9602114.}
\REFSCON\MVII{D.~R.~Morrison and C.~Vafa, Nucl. Phys. {\bf B476} (1996) 437, 
hep-th/9603161.}
\REFSCON\Ber{M.~Bershadsky \etal, hep-th/9605200.} 
\refsend).
It was observed in \cite{\CF}~that it is the dual polyhedron, $\nabla$, which
exhibits a regular structure which makes it possible, in particular, to 
determine the enhanced gauge symmetry given~$\nabla$. It was noted also that 
in all the examples of heterotic/type II dual pairs the $K3$ and elliptic
fibration structure shows itself in the existence of three- and two-dimensional
reflexive subpolyhedra, respectively, inside~$\nabla$.
The three-dimensional reflexive subpolyhedron which corresponds
to the generic $K3$ fiber was shown to contain the information
about the part of the total gauge group (the only part in the examples 
considered in \cite{\CF}) which has perturbative interpretation on the 
heterotic side. These results were extended to include non-perturbative gauge
groups in~
\Ref\BG{P.~Candelas, E.~Perevalov and G.~Rajesh, hep-th/9704097.}.

The purpose of the present paper is to extend this dictionary to
include the charged matter content of the low energy effective theories.
The question of determining the charged matter content from geometry was
addressed in Refs.~
\REFS\KV{S.~Katz and C.~Vafa, hep-th/9606086.}
\REFSCON\KKV{S.~Katz, A.~Klemm and C.~Vafa, hep-th/9609239.}
\REFSCON\BKKV{P.~Berglund, S.~Katz, A.~Klemm and P.~Mayr, hep-th/9605154.}
\REFSCON\AG{P.~S.~Aspinwall and M.~Gross, hep-th/9605131.}
\REFSCON\A{P.~S.~Aspinwall, hep-th/9611137.}
\refsend , where it was pointed out that gauge groups are associated with
curves of singularities and that intersecting curves of singularities lead to
matter charged under both gauge groups. In this paper we use the
requirement of anomaly cancellation~
\Ref\Sadov{V.~Sadov, hep-th/9606008.}\ to relate the charged matter content
directly
to the toric data. The duality \eqref{duality} applies most directly to \cys\
that
are elliptic fibrations. for these manifolds the charged matter is associated
with
the divisors of the base, $B$, of the fibration and the number and group
representation of these fields is determined by the intersection numbers of the
divisors.

The organisation of this paper is as follows. In \SS{2}, we list relevant
results in toric geometry. In \SS{3}, we describe the relation between the
charged matter content and intersection theory, and work out some simple
examples. \SS{4} is devoted to applying this technique to some of the models of
Ref.~\cite{\BG}. \SS{5} summarises our results.
\newpage
\section{Toric}{Some Results in Toric Geometry}
We will be dealing with elliptic \cyt s which are described as hypersurfaces
in toric varieties. Our notation and conventions follow those of our companion
paper\cite{\BG}. We shall briefly review the relevant results so that
the present is self-contained. The reader, however, may care to refer to
\cite{\BG} for a fuller account. 

In the following, $\L$ is a lattice of rank $n$ and $\D$, the Newton
Polyhedron of the \cy\ $n$-fold, is a reflexive polyhedron in $\L$.
We denote by $\D$ the polyhedron dual to $\D$ and by $V$ the lattice dual
to $\L$. The real extension of $V$ is denoted by $V_{\IR}$.
It has been shown in~
\Ref\AKMS{A.~Avram, M.~Kreuzer, M.~Mandelberg
and H.~Skarke, hep-th/9610154.}~that in order for a \cy\ $n$-fold to be
a fibration with generic fiber a \cy\ $(n-k)$-fold, it is necessary and 
sufficient that 
\item\ {(i) There is a projection operator $\P$: $\L\rightarrow \L_{n-k}$,
where 
$\L_{n-k}$ is an $n-k$ dimensional sublattice, such that $\P(\D)$ is a
reflexive polyhedron in $\L_{n-k}$, or}
\item\ {(ii) There is a lattice plane, $h$, in $V_{\IR}$ through the origin 
whose intersection with $\nabla$ is an $n-k$ dimensional reflexive polyhedron,
{\it i.e.\/} it is a slice of the polyhedron.}

\noindent (i) and (ii) are equivalent conditions. In case (i) the
polyhedron of the fiber appears
as a {\sl projection\/} while in case (ii) it appears as an
{\sl injection\/}, the projection and the injection being related by mirror
symmetry.
In particular, if the polyhedron of the $(n-k)$-dimensional \cym\ exists as
both a projection and an injection, then the intersection in $\nabla$
is also a certain projection implying that the mirror manifold is a fibration
with an $n-k$ dimensional \cym\ as the typical fiber.
If (i) or (ii) hold there is also a way to to see the base of the fibration
torically~\REFS\rSp{M.~Kreuzer and H.~Skarke, hep-th/9701175.}\refsend
.
The hyperplane $h$ generates a $n-k$ dimensional sublattice of $V$. Denote this
lattice $V_{\rm fiber}$. Then the quotient lattice 
$V_{\rm base}=V/V_{\rm fiber}$ is the lattice in which the fan of the base 
lives. The fan itself can be constructed as follows. Let $\P_B$ be a projection
operator acting in $V$, of rank dim$(V)-2$, such that it projects $h$ onto
a point. Then
$\P_B(V)=V_{\rm base}$. When $\P_B$ acts on $\nabla$ the result is a $k$ 
dimensional
set of points in $V_{\rm base}$ which gives us the fan of the base if we draw
rays through each point in the set. The pre-image of every ray in the base
under $\P_B$ determines the type of singularity (including the monodromy, if
any) along the corresponding curve in the base in the way described
in~\cite{\BG}. Thus, each ray is associated with a factor (which may be
trivial) of the total gauge group.

For elliptic \cyt s which are hypersurfaces in toric varieties, the base of
the fibration is a nonsingular two-dimensional toric variety.
These are well decribed in \SS{2.5} of Ref.~
\Ref\Ful{W.~Fulton, Introduction to Toric Varieties, Princeton University
Press, 1993.} and are specified by giving a sequence of lattice points
$$ v_0, v_1, \ldots, v_{l-1}, v_d = v_0 $$
in counterclockwise order, in $V = \IZ^{2}$, such that the $v_i$'s are the
first lattice points in each ray, and successive pairs
generate the lattice (see Figure~\figref{rays}). In general,
these satisfy
$$ a_iv_i = v_{i-1} + v_{i+1},\;\;  1\leq i \leq d , $$ 
for some integers $a_i$.
\midinsert
\def\rays{
\vbox{\vskip10pt\hbox{\epsfxsize=2.0truein\epsfbox{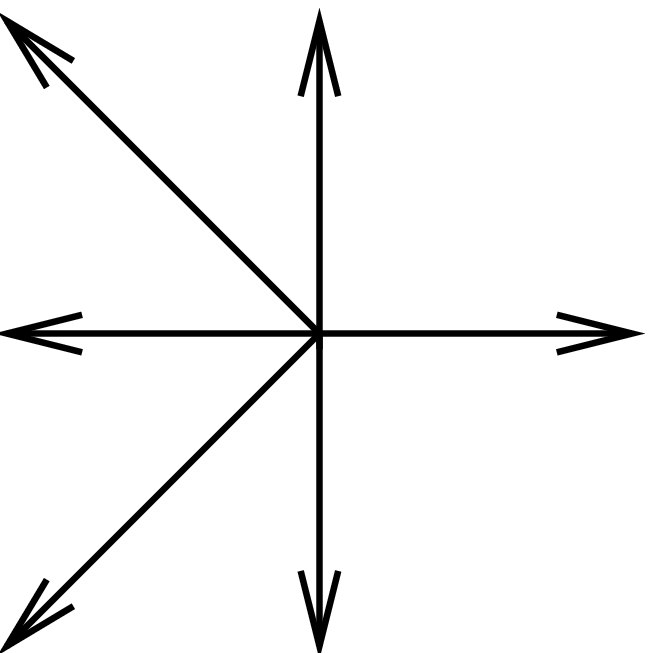}}}}
\figbox{\rays\vskip0pt}{\figlabel{rays}}
{The fan of a typical two-dimensional toric variety.}
\place{3.4}{2.75}{$v_1$}
\place{2.05}{2.75}{$v_2$}
\place{4.3}{1.9}{$v_0 = v_6$}
\place{2.05}{0.9}{$v_4$}
\place{2.05}{1.9}{$v_3$}
\place{3.4}{0.9}{$v_5$}
\endinsert

These two dimensional surfaces are readily classified as follows. For $d=3$,
the surface is
$\IP_{2}$, for $d=4$, one gets a Hirzebruch surface $\IF_{n}$ for some $n$.
All higher
values of $d$ yield surfaces which are obtained by successive blow-ups of
either $\IP_{2}$ or $\IF_{n}$ at fixed points of the torus action.

Each $v_i$ determines a curve $D_i \cong \IP_{1}$ in the variety. The normal
bundle to this embedding is the line bundle \ca{O}($-a_i$) on $\IP_{1}$.
Successive curves meet transversally, but are otherwise disjoint:
$$(D_i\cdot D_j) = \cases{1 &if $\vert i - j\vert = 1$;\cr\noalign{\vskip2pt}
                         -a_i &if $i = j$;\cr\noalign{\vskip2pt}
                         0 &otherwise.\cr\noalign{\vskip2pt}}$$
Finally, the canonical class of the surface is given by
$$K = - \sum_i D_i ,$$
where the sum runs over all the divisors corresponding to the lattice points
$v_i$ in the fan of the base.       
\section{matter}{Identifying the Matter Content}
In this section, we describe our approach to determining the charged matter
content from the toric data.
On the heterotic side, the matter content of the perturbative vacua can be
found by applying the index theorem.
Let $H_{1,2}$ be the background gauge groups (simple  
subgroups of $E_8$) and $k_{1,2}$ the corresponding instanton numbers (second
Chern classes of the background gauge bundles on the $K3$). The contribution
of each $E_8$ to the unbroken gauge group in six dimensions is then the 
commutant $G_{1,2}$ of $H_{1,2}$ respectively. The number of hypermultiplets 
in the representation $R_a$ of $G$ is then
$$N(R_a)=kT(M_a)-{\rm dim}(M_a),$$
where the adjoint of $E_8$ decomposes under $G\times H$ as {\bf 248}=$\sum_{a}
(R_a, M_a)$, and $T(M_a)$ is given by
tr$(T_{a}^{i}T_{a}^{j})=T(M_a)\delta _{ij}$, 
$T_{a}^{i}$ being a generator of $H$ in the representation $M_a$. 
\midinsert
$$\vbox{\offinterlineskip\halign{\strut # height 12pt depth 5pt
&\hfil\quad $#$\quad\hfil\vrule
&\hfil\quad $#$\quad\hfil\vrule \cr
\noalign{\hrule}
\omit{\vrule height3pt}&&\cr
\vrule&$Group$ 
       &$Charged Matter Content$ \cr
\omit{\vrule height3pt}&&\cr
\noalign{\hrule\vskip3pt\hrule}
\omit{\vrule height2pt}&&\cr
\vrule&SU(2)&(6n+16){\bf 2}\cr
\vrule&SU(2)_2&(8n+32){\bf 2} + (n-1){\bf 3}\cr
\vrule&SU(3)&(6n+18){\bf 3}\cr
\vrule&SO(5)&(n+1){\bf 5}+(4n+16){\bf 4}\cr
\vrule&G_2&(3n+10){\bf 7}\cr
\vrule&SU(4)&(n+2){\bf 6}+(4n+16){\bf 4}\cr
\vrule&SO(7)&(n+3){\bf 7}+(2n+8){\bf 8}\cr
\vrule&Sp(3)&(16+2n+{3\over 2}r){\bf 6}+(n+1-r){\bf 14}+{1\over 2}r{\bf 14}'\cr
\vrule&SO(8)&(n+4)({\bf 8}_c+{\bf 8}_s+{\bf 8}_v)\cr
\vrule&SU(5)&(3n+16){\bf 5}+(2+n){\bf 10}\cr
\vrule&SO(9)&(n+5){\bf 9}+(n+4){\bf 16}\cr
\vrule&F_4&(n+5){\bf 26}\cr
\vrule&SU(6)&{r\over 2}{\bf 20}+(16+r+2n){\bf 6}+(2+n-r){\bf 15}\cr
\vrule&SO(10)&(n+4){\bf 16}+(n+6){\bf 10}\cr
\vrule&SO(11)&({n\over 2}+2){\bf 32}+(n+7){\bf 11}\cr
\vrule&SO(12)&{r\over 2}{\bf 32}+({{4+n-r}\over 2}){\bf 32}'+(n+8){\bf 12}\cr
\vrule&E_6&(n+6){\bf 27}\cr
\vrule&E_7&({n\over 2}+4){\bf 56}\cr
\omit{\vrule height2pt}&&\cr
\noalign{\hrule}
}}
$$
\nobreak\tablecaption{charged}{Charged matter content of models with enhanced
gauge symmetry, as a function of $n$.}
\endinsert

Anomaly cancellation requires $k_1+k_2=24$, so it is convenient to define
$$n=k_1-12=12-k_2$$ and take $n\ge 0$ (\ie $k_1\ge k_2$). We find that the
massless spectrum satisfies $H - V = 244$,where $H$ and $V$ are the numbers of
massless hypermultiplets and vector multiplets, respectively.
If $n\le 8$ we can
take $H_1=H_2=SU(2)$ and
obtain $E_7\times E_7$ gauge symmetry in six dimensions
with the following matter content:
$${1\over 2}(8+n)({\bf 56}, {\bf 1})+{1\over 2}(8-n)({\bf 1}, {\bf 56})
+ 62({\bf 1}, {\bf 1})~.$$
If $9\le n\le 12$, then $k_2$ cannot support an $SU(2)$ background, and the 
instantons in the second $E_8$ are necessarily small producing an unbroken
$E_8$. The gauge group in six dimensions is thus $E_7\times E_8$ with matter
content
$${1\over 2}(8+n)({\bf 56}, {\bf 1})+(53+n)({\bf 1}, {\bf 1})~.$$
Models with subgroups of the above can be obtained by gauge symmetry 
breaking via Higgs mechanism, or, equivalently, by taking the subgroups of
$E_8$ other than $SU(2)$ as~$H_{1,2}$.

Models with additional tensor multiplets, corresponding to non-perturbative
heterotic vacua also exist. For these models, the massless spectrum satisfies
$$ H - V = 273 - 29T \eqlabel{anomaly}$$
where $T$ is the number of massless tensor multiplets (perturbative heterotic
models have $T=1$).
We can now determine the matter content for any unbroken
group. We list some of the results in Table~\tabref{charged}, which appeared
in Ref.~\cite{\Ber}, where it was observed that the charged matter content is
encoded on the F-theory side in the degree of vanishing of the discriminant on
the locus of the corresponding singularity.

It was shown in~\cite{\Sadov} that the Green-Schwarz anomaly cancellation
condition puts restrictions on the matter content of the six-dimensional gauge
theories obtained on compactifying F-theory on an elliptic \cyt .
In particular, the amount of matter charged with respect
to a group corresponding to a divisor $D_i$ in the base of the elliptic
fibration satisfies
$$\eqalign{& \hbox{index}(Ad_i) - \sum_R \hbox{index}(R_i)n_{R_i} =
6(K\cdot D_i)\cr
           & y_{Ad_i} - \sum_R y_{Ad_i}n_{R_i} = - 3 (D_i\cdot D_i),}
\eqlabel{totalmatter}$$
where in this expression $n_{R_i}$ denotes the total number of hypermultiplets
in the representation
$R_i$ of the group $G_i$ and $Ad_i$ denotes the adjoint representation. The
index($R_i$) is
given by trace$(T^a_iT^b_i) = \hbox{index}(R_i)\delta^{ij}$ and $y_{R_i}$ is
defined by decomposing $\hbox{tr}_{R_i}F^4 = x_{R_a}\hbox{tr}F^4 +
y_{R_a}(\hbox{tr}F^2)^2$,
assuming $R_i$ has two independent fourth order invariants, and tr is the
trace in a preferred representation, which is the fundamental for $SU(n)$.
If $R_a$ has only one fourth order invariant, then $x_{R_a} = 0$.  

In addition, matter charged with respect to two groups satisfies
$$\sum_{R,R'} \hbox{index}(R_a)\,\hbox{index}(R_b')\,n_{R_aR_b'} =
(D_a\cdot D_b).
\eqlabel{multmatter}$$

It is easy to derive the matter content for the perturbative
(on the heterotic side) gauge groups
listed in Table~\tabref{charged} using these formulae. Note that the value of
$n$ in the table is precisely the self-intersection of the corresponding
divisor in the base.
We can now also obtain the matter content for the groups which have a
non-perturbative origin on the heterotic side, since the formulae above do not
depend upon whether the groups are perturbative or non-perturbative from the
heterotic point of view.    

We illustrate this approach with a few simple examples. Consider the 
${\rm Spin}(32)/\IZ_2$ heterotic string on a $K3$ manifold with one instanton 
shrunk to a point~
\Ref\Wsi{E. Witten, \npb 460 (1996) 541, hep-th/9511030.}.
The corresponding F-theory dual is an elliptic fibration over $F_4$, whose fan
is shown in Fig.~\figref{fan}.
There is an $I_1^{\ast}$ singularity along the zero section $C_0$ of $F_4$ 
(which is a $\IP_1$ bundle over $\IP_1$), with a monodromy action on the 
degenerate fiber leading to $SO(9)$ gauge group as well as $I_2$ singularity
along the divisor corresponding to $\IP_1$ fiber, $f$, leading to an $SU(2)$
gauge group. As we know (and as is easily found from the fan of $F_4$) the
intersection numbers in this case are as follows:
$$\eqalign{&C_0\cdot C_0=-4\cr
          &f\cdot f=0\cr
          &C_0\cdot f=1\cr}$$
Thus, using Eq.~\eqref{totalmatter}, we find that there
are $-4+5=1$ {\bf 9}'s of $SO(9)$
and $6\cdot 0+16=16$ {\bf 2}'s of $SU(2)$ in the hypermultiplet spectrum.
Using Eq.~\eqref{multmatter}, we find that the charged matter content is
${1\over 2}({\bf 9}, {\bf 2})+{23\over 2}({\bf 1}, {\bf 2})$, which agrees
with the known result.

If we shrink two instantons at the same point of the $K3$, the $SO(9)$ of the
previous 
example becomes $SO(10)$, and the $SU(2)$ becomes $Sp(2)\cong SO(5)$. The 
intersection numbers are unchanged. Applying Eq.~\eqref{totalmatter} we
obtain $-4+6=2$ 
{\bf 10}'s of $SO(10)$ and $0+1=1$ {\bf 5}'s as well as $4\cdot 0+16=16$
{\bf 4}'s of $Sp(2)$. Eq.~\eqref{multmatter} now fixes the matter content to be
${1\over 2}({\bf 10}, {\bf 4})+({\bf 1}, {\bf 5})+11({\bf 1}, {\bf 4})$,
which again agrees with the known result.
\midinsert
\def\fan{
\vbox{\vskip10pt\hbox{\epsfxsize=1.5truein\epsfbox{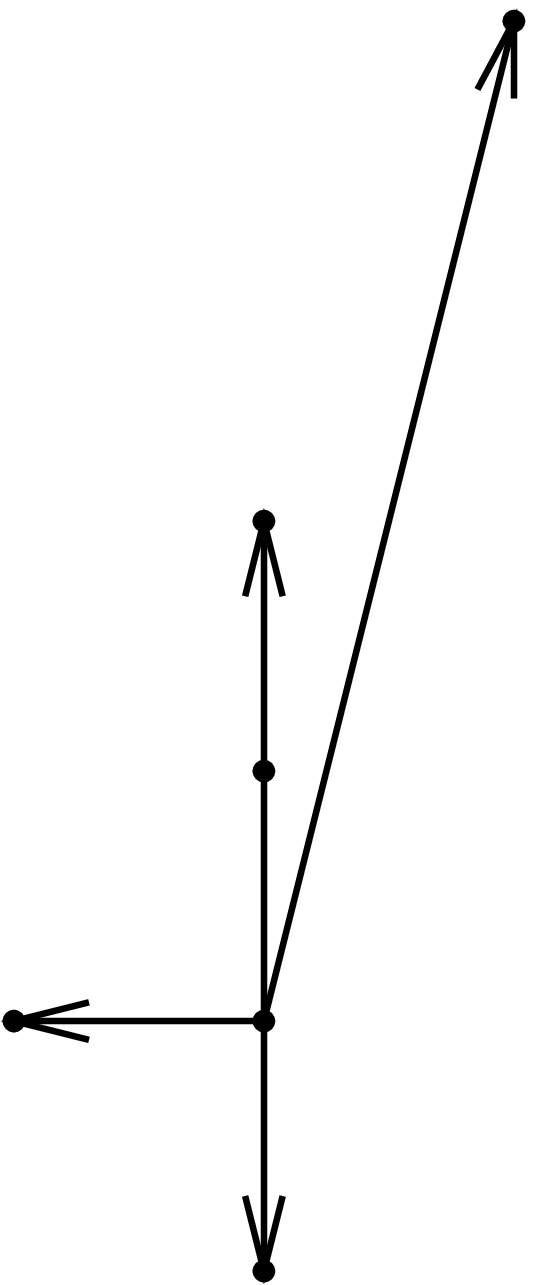}}}}
\figbox{\fan\vskip0pt}{\figlabel{fan}}
{The fan of $\IF_4$. The rays of the fan are labelled by the divisors to
which they correspond.}
\place{2.8}{1.1}{(0,-1)}
\place{3.4}{1.1}{$C_0 + 4f$}
\place{3.4}{1.6}{(0,0)}
\place{3.75}{3.1}{$f$}
\place{4.1}{4.5}{(1,4)}
\place{3.0}{2.4}{$C_0$}
\place{3.2}{3.3}{(0,2)}
\place{2.75}{1.5}{$f$}
\place{2.1}{1.6}{(-1,0)}
\endinsert
\newpage
\section{eg}{Examples}
In this section, we apply our method to some of the examples of
Ref.~\cite{\BG}. We have chosen to study the matter content of only those
models for which the gauge content is known with certainty, \ie those for
which the number of non-toric corrections is zero.
\subsection{The mirrors of models with $n=0$}
\subsubsection{(i)}{The mirror of the manifold with enhanced gauge group
SU(1)} 
This model has gauge group
$E_8^8{\times} F_4^8{\times} G_2^{16}{\times} SU(2)^{16}$ and 97 tensors.
Analysis of the polyhedron reveals that all the divisors corresponding to
the $E_8$ factors have self-intersection $-12$, those corresponding to the
$F_4$'s have self-intersection $-5$, those corresponding to the $G_2$'s have
self-intersection $-3$, and those corresponding to the $SU(2)$'s all have
self-intersection $-2$. Furthermore, the only divisors among these that
intersect each other are those corresponding to the $G_2$'s and the $SU(2)$'s,
which intersect each other pairwise, \ie every $G_2$ divisor intersects
exactly one divisor, which corresponds to an $SU(2)$, and vice versa.
{}From the self-intersections, we conclude that the $E_8$'s and $F_4$'s are all
matter-free, while each $G_2$ comes with one {\bf 7}, and each $SU(2)$ comes
with four {\bf 2}'s. Since each $G_2$ intersects an $SU(2)$, we conclude that
the charged matter consists of $\{{1\over 2}({\bf 7},
{\bf 2})+{1\over 2}({\bf 1}, {\bf 2})\}$ from each of the 16 factors of
$G_2{\times} SU(2)$\ for a total of 128 charged hypermultiplets.
Thus, for this vacuum, $H = H_c + H_0 = 128 + 4 = 132$,
$V= \hbox{dim}(G) = 2672$,
$$H - V = 132 - 2672 = -2540 = 273 - 29{\times} 97 = 273 - 29T,$$
so that the anomaly
cancellation condition~\eqref{anomaly} is satisfied. We find that this is true
in all the cases that we have studied, and, in particular, for the examples
below. This result is no surprise, since we obtain the matter content from the
Green-Schwarz anomaly cancellation condition.
\medskip

\subsubsection{(ii)}{The mirror of the manifold with enhanced gauge group
SU(2)} 
The gauge group is
$E_8^5{\times} E_7^3{\times} F_4^6{\times} G_2^{12}{\times} SO(7)^2{\times}
SU(2)^{16}$
and, in addition, there are 81 tensors. We find that the self-intersections of
the divisors are such that the $E_8$'s, $E_7$'s and $F_4$'s are all
matter-free. Also we find 12
pairs of mutually intersecting $G_2$ and $SU(2)$ divisors each of which have
self-intersections $-3$ and $-2$ respectively. Here, and in future, we adopt
the following shorthand for mutually intersecting divisors: we list the group
factors in the form $G_1^{[a_1]}{\times} G_2^{[a_2]}{\times} \ldots
G_n^{[a_n]}$,
so that the corresponding divisors have the following intersection pattern
$$(D_i\cdot D_j) = \cases{1 &if $\vert i - j\vert = 1$;\cr\noalign{\vskip2pt}
                         -a_i &if $i = j$;\cr\noalign{\vskip2pt}
                         0 &otherwise.\cr\noalign{\vskip2pt}}$$     
\noindent where $D_i$ is the divisor corresponding to group factor $G_i$.
Thus we would say that there are 12 factors of $G_2^{[-3]}{\times}
SU(2)^{[-2]}$.
We thus obtain 12 sets of 
$$\{{1\over 2}({\bf 7}, {\bf 2})+{1\over 2}({\bf 1}, {\bf 2})\}.$$
\noindent In addition, we get 2 factors of
$SU(2)^{[-2]}{\times} SO(7)^{[-3]}{\times} SU(2)^{[-2]}$, so we find 2 sets of
$$\{{1\over 2}({\bf 2}, {\bf 8}, {\bf 1})+{1\over 2}({\bf 1}, {\bf 8},
{\bf 2})\}$$
\noindent for a total of 128 charged hypermultiplets. The combination
$G_2^{[-3]}{\times} SU(2)^{[-2]}$ occurs repeatedly in each of the examples
below and in each case contributes
$\{{1\over 2}({\bf 7}, {\bf 2})+{1\over 2}({\bf 1}, {\bf 2})\}$. 
\medskip

\subsubsection{(iii)}{The mirror of the manifold with enhanced gauge group
SU(3)}
The gauge group is $E_8^5{\times} E_6^3{\times} F_4^6{\times}
G_2^{10}{\times} SU(3)^4{\times} SU(2)^{10}$
and, in addition, there are 75 tensors. The only groups that contain charged
matter are 10 factors of $G_2^{[-3]}{\times} SU(2)^{[-2]}$ for a total of 80
charged hypermultiplets.
\medskip

\subsubsection{(iv)}{The mirror of the manifold with enhanced gauge group
SU(2)$^2$}
The group is 
$E_8^5{\times} F_4^4{\times} G_2^{10}{\times} SO(5)^2{\times} SO(9)^2{\times}
SO(11)^2{\times} SO(12){\times} SU(2)^{12}$, and there are 67
tensors. The charged matter content comes from 10 factors of
$G_2^{[-3]}{\times} SU(2)^{[-2]}$ plus
matter charged under $SO(9)^{[-4]} \times SU(2)^{[-1]} \times
SO(11)^{[-4]} \times SO(5)^{[-1]} \times
SO(12)^{[-4]} \times $\break$ SO(5)^{[-1]} \times SO(11)^{[-4]} \times
SU(2)^{[-1]} \times SO(9)^{[-4]}$
as follows:
$$\def\skip{\hphantom{\bf 1}} 
\eqalign{\{ & {1\over 2}(\skip {\bf 9},\skip {\bf 2},\skip {\bf 1}, 
\skip {\bf 1},\skip {\bf 1},\ldots,
\skip {\bf 1},\skip {\bf 1})\cr
           + & {1\over 2}(\skip {\bf 1}, \skip {\bf 2}, {\bf 11}, 
\skip {\bf 1}, \skip {\bf 1}, \ldots,
\skip {\bf 1}, \skip {\bf 1})\cr 
           +& {1\over 2}(\skip {\bf 1}, \skip {\bf 1}, {\bf 11},
\skip {\bf 4}, \skip {\bf 1}, \ldots,
\skip {\bf 1}, \skip {\bf 1})}$$
$$\def\skip{\hphantom{\bf 1}} 
\eqalign{ +& {1\over 2}(\skip {\bf 1}, \skip {\bf 1}, \skip {\bf 1},
\skip {\bf 4}, \skip {\bf 1}, \ldots,
\skip {\bf 1}, \skip {\bf 1})\cr 
           +& {1\over 2}(\skip {\bf 1}, \skip {\bf 1}, \skip {\bf 1},
\skip {\bf 4}, {\bf 12},
\skip {\bf 1}, \ldots, \skip {\bf 1})\cr
            & \vdots \cr
           +& {1\over 2}(\skip {\bf 1}, \skip {\bf 1}, \skip {\bf 1},
\skip {\bf 1}, \skip {\bf 1}, \ldots,
\skip {\bf 2}, \skip {\bf 9})\}}$$  
\noindent for a total of 216 charged hypermultiplets.
\medskip

\subsubsection{(v)}{The mirror of the manifold with enhanced gauge group SU(4)}
The gauge group is
$E_8^5{\times} F_4^4{\times} G_2^{10}{\times} SO(9)^2{\times} SO(10)^3{\times}
SU(2)^{14}$,
and there are 67 tensors.
The charged matter comes from 10 factors of $G_2^{[-3]}{\times} SU(2)^{[-2]}$
plus matter charged under 
$SO(9)^{[-4]} \times SU(2)^{[-1]} \times
SO(10)^{[-4]} \times SU(2)^{[-1]} \times
SO(10)^{[-4]} \times SU(2)^{[-1]} \times SO(10)^{[-4]} \times $\break$
SU(2)^{[-1]} \times SO(9)^{[-4]}$  as follows:
$$\def\skip{\hphantom{\bf 1}}
\eqalign{\{& {1\over 2}(\skip {\bf 9}, \skip {\bf 2}, \skip {\bf 1}, \skip
{\bf 1}, \skip {\bf 1}, \ldots,
\skip {\bf 1}, \skip {\bf 1})\cr
           +& {1\over 2}( \skip {\bf 1}, \skip {\bf 2}, \skip {\bf 1}, \skip
{\bf 1}, \skip {\bf 1}, \ldots,
\skip {\bf 1}, \skip {\bf 1})\cr 
           +& {1\over 2}(\skip {\bf 1}, \skip {\bf 2}, {\bf 10}, \skip {\bf 1},
\skip  {\bf 1}, \ldots,
\skip {\bf 1}, \skip {\bf 1})\cr
           +& {1\over 2}( \skip {\bf 1}, \skip {\bf 1}, {\bf 10}, \skip
{\bf 2}, \skip {\bf 1}, \ldots,
\skip {\bf 1}, \skip {\bf 1})\cr 
           +& {1\over 2}(\skip {\bf 1}, \skip {\bf 1}, \skip {\bf 1}, \skip
{\bf 2}, {\bf 10},
\skip {\bf 1}, \ldots, \skip {\bf 1})\cr
           & \vdots \cr
          +&{1\over 2}(\skip {\bf 1}, \skip {\bf 1}, \skip {\bf 1}, \skip
{\bf 1}, \skip {\bf 1}, \ldots,
\skip {\bf 2}, \skip {\bf 9})\}}$$
for a total of 160 charged hypermultiplets.
Note that this model can be obtained by the following Higgsing of the gauge
group of the previous model:
$$ SO(5)\rightarrow SU(2),\; SO(12)\rightarrow SO(10),\; \hbox{and} \;
SO(11)\rightarrow SO(10).$$
It is easy to verify that this breaking produces the charged matter content
above, and precisely one additional singlet, which corresponds to the
difference in the ranks of $SU(4)$ and $SU(2)^2$.
\medskip

\subsubsection{(vi)}{The mirror of the manifold with enhanced gauge group
SU(2)${\times}$ SU(3)}
The gauge group is $E_8^5{\times} F_4^4{\times} G_2^{8}{\times} SU(6){\times}
SU(5)^2{\times} SU(4)^2{\times} SU(3)^2{\times} SU(2)^{10}$, and there are 61
tensors. 
The charged matter comes from 8 factors of $G_2^{[-3]}{\times} SU(2)^{[-2]}$
plus matter that is charged under $SU(2)^{[-2]}{\times} SU(3)^{[-2]}{\times}
SU(4)^{[-2]}{\times}
SU(5)^{[-2]}{\times} SU(6)^{[-2]}{\times} SU(5)^{[-2]}$\break${\times} 
SU(4)^{[-2]}{\times} SU(3)^{[-2]}{\times} SU(2)^{[-2]}$ as
follows:
$$\def\skip{\hphantom{\bf 1}}
\eqalign{\{& ( \skip {\bf 2}, \skip {\bf 1}, \skip {\bf 1}, \skip {\bf 1},
\skip {\bf 1}, \ldots,
\skip {\bf 1}, \skip {\bf 1})\cr
           +& ( \skip {\bf 2}, \skip {\bf 3}, \skip {\bf 1}, \skip {\bf 1},
\skip {\bf 1}, \ldots,
\skip {\bf 1}, \skip {\bf 1})\cr 
           +& ( \skip {\bf 1}, \skip {\bf 3}, \skip {\bf 4}, \skip {\bf 1},
\skip {\bf 1}, \ldots,
\skip {\bf 1}, \skip {\bf 1})\cr
           +& (\skip {\bf 1}, \skip {\bf 1}, \skip {\bf 4}, \skip {\bf 5},
\skip {\bf 1}, \ldots,
\skip {\bf 1}, \skip {\bf 1})\cr 
           +& (\skip {\bf 1}, \skip {\bf 1}, \skip {\bf 1}, \skip {\bf 5},
\skip {\bf 6},
\skip {\bf 1}, \ldots, \skip {\bf 1})\cr
          +2& ( \skip {\bf 1}, \skip {\bf 1}, \skip {\bf 1}, \skip {\bf 1},
\skip {\bf 6},
\skip {\bf 1}, \ldots, \skip {\bf 1})\cr
           & \vdots \cr
           +& ( \skip {\bf 1}, \skip {\bf 1}, \skip {\bf 1}, \skip {\bf 1},
\skip {\bf 1}, \ldots,
\skip {\bf 1}, \skip {\bf 2})\}}$$   
for a total of 216 charged hypermultiplets.
\medskip

\subsubsection{(vii)}{The mirror of the manifold with enhanced gauge group
SU(5)}
The gauge group is $E_8^5{{\times}} F_4^4{\times} G_2^{8}{\times}
SU(5)^3{\times} SU(4)^2{\times} SU(3)^2{\times} SU(2)^{10}$, and there are 61
tensors.
The charged matter comes from 8 factors of $G_2^{[-3]}{\times} SU(2)^{[-2]}$
plus matter that is charged under $SU(2)^{[-2]} \times SU(3)^{[-2]} \times
SU(4)^{[-2]} \times SU(5)^{[-2]} \times SU(5)^{[-2]} \times
SU(5)^{[-2]} \times $\break$ SU(4)^{[-2]} \times SU(3)^{[-2]} \times
SU(2)^{[-2]}$ 
as follows:
$$\def\skip{\hphantom{\bf 1}}
\eqalign{\{& ( \skip {\bf 2}, \skip {\bf 1}, \skip {\bf 1}, \skip {\bf 1},
\skip {\bf 1}, \ldots,
\skip {\bf 1}, \skip {\bf 1})\cr
           +& ( \skip {\bf 2}, \skip {\bf 3}, \skip {\bf 1}, \skip {\bf 1},
\skip {\bf 1}, \ldots,
\skip {\bf 1}, \skip {\bf 1})\cr 
           +& ( \skip {\bf 1}, \skip {\bf 3}, \skip {\bf 4}, \skip {\bf 1},
\skip {\bf 1}, \ldots,
\skip {\bf 1}, \skip {\bf 1})\cr
           +& ( \skip {\bf 1}, \skip {\bf 1}, \skip {\bf 4}, \skip {\bf 5},
\skip {\bf 1}, \ldots,
\skip {\bf 1}, \skip {\bf 1})\cr 
           +& ( \skip {\bf 1}, \skip {\bf 1}, \skip {\bf 1}, \skip {\bf 5},
\skip {\bf 1},
\skip {\bf 1}, \ldots, \skip {\bf 1})\cr
           +& ( \skip {\bf 1}, \skip {\bf 1}, \skip {\bf 1}, \skip {\bf 5},
\skip {\bf 5},
\skip {\bf 1}, \ldots, \skip {\bf 1})\cr
           & \vdots \cr
           +&( \skip {\bf 1}, \skip {\bf 1}, \skip {\bf 1}, \skip {\bf 1},
\skip {\bf 1}, \ldots,
\skip {\bf 1}, \skip {\bf 2})\}}$$   
for a total of 204 charged hypermultiplets.
It is easy to see that this matter content can be obtained from the previous
one by Higgsing the $SU(6)$ to
$SU(5)$, yielding exactly one extra singlet, which corresponds to the
difference between the ranks of $SU(5)$ and $SU(2){\times} SU(3)$.
\medskip

\subsubsection{(viii)}{The mirror of the manifold with enhanced gauge group
SO(10)}
The gauge group is $E_8^5{\times} F_4^4{\times} G_2^{8}{\times}
SU(4)^5{\times} SU(3)^2{\times} SU(2)^{10}$, and there are 61
tensors.
The charged matter comes from 8 factors of $G_2^{[-3]}{\times} SU(2)^{[-2]}$
plus matter that is charged under 
$SU(2)^{[-2]} \times SU(3)^{[-2]} \times SU(4)^{[-2]} \times
SU(4)^{[-2]} \times SU(4)^{[-2]} \times SU(4)^{[-2]} \times
SU(4)^{[-2]} \times $\break$ SU(3)^{[-2]} \times SU(2)^{[-2]}$ as
follows:
$$\def\skip{\hphantom{\bf 1}}
\eqalign{\{& ( \skip {\bf 2}, \skip {\bf 1}, \skip {\bf 1}, \skip {\bf 1},
\skip {\bf 1}, \ldots, \skip {\bf 1}, \skip {\bf 1})\cr
           +& ( \skip {\bf 2}, \skip {\bf 3}, \skip {\bf 1}, \skip {\bf 1},
\skip {\bf 1}, \ldots, \skip {\bf 1}, \skip {\bf 1})\cr 
           +& (\skip {\bf 1}, \skip {\bf 3}, \skip {\bf 4}, \skip {\bf 1},
\skip {\bf 1}, \ldots, \skip {\bf 1}, \skip {\bf 1})\cr
           +& ( \skip {\bf 1}, \skip {\bf 1}, \skip {\bf 4}, \skip {\bf 1},
\skip {\bf 1}, \ldots, \skip {\bf 1}, \skip {\bf 1})\cr 
           +& ( \skip {\bf 1}, \skip {\bf 1}, \skip {\bf 4}, \skip {\bf 4},
\skip {\bf 1}, \skip {\bf 1}, \ldots, \skip {\bf 1})\cr
           +& ( \skip {\bf 1}, \skip {\bf 1}, \skip {\bf 1}, \skip {\bf 4},
\skip {\bf 4}, \skip {\bf 1}, \ldots, \skip {\bf 1})\cr
            & \vdots \cr
           +& ( \skip {\bf 1}, \skip {\bf 1}, \skip {\bf 1}, \skip {\bf 1},
\skip {\bf 1}, \ldots, \skip {\bf 1}, \skip {\bf 2})\}}$$
for a total of 176 charged hypermultiplets.
Again, one finds that this matter content can be obtained from the
previous one by Higgsing the $SU(5)$'s to $SU(4)$'s, yielding exactly one
extra singlet, corresponding to the difference between the ranks of
$SO(10)$ and $SU(5)$.
\medskip

\subsubsection{(ix)}{The mirror of the manifold with enhanced gauge group
E$_6$}
The gauge group is $E_8^5{\times} F_4^4{\times} G_2^{8}{\times}
SU(3)^7{\times} SU(2)^{10}$, and there are 61
tensors. The charged matter comes from 8 factors of $G_2^{[-3]}{\times}
SU(2)^{[-2]}$
plus matter that is charged under\break
$SU(2)^{[-2]} \times SU(3)^{[-2]} \times SU(3)^{[-2]} \times
SU(3)^{[-2]} \times SU(3)^{[-2]} \times SU(3)^{[-2]} \times
SU(3)^{[-2]} \times $\break$ SU(3)^{[-2]} \times  SU(2)^{[-2]}$ as
follows:
$$\def\skip{\hphantom{\bf 1}}
\eqalign{\{& ( \skip {\bf 2}, \skip {\bf 1}, \skip {\bf 1}, \skip {\bf 1},
\skip {\bf 1}, \ldots,
\skip {\bf 1}, \skip {\bf 1})\cr
           +& ( \skip {\bf 2}, \skip {\bf 3}, \skip {\bf 1}, \skip {\bf 1},
\skip {\bf 1}, \ldots,
\skip {\bf 1}, \skip {\bf 1})\cr 
           +& (\skip {\bf 1}, \skip {\bf 3}, \skip {\bf 1}, \skip {\bf 1},
\skip {\bf 1}, \ldots,
\skip {\bf 1}, \skip {\bf 1})\cr
           +& ( \skip {\bf 1}, \skip {\bf 3}, \skip {\bf 3}, \skip {\bf 1},
\skip {\bf 1}, \ldots,
\skip {\bf 1}, \skip {\bf 1})\cr 
           +& ( \skip {\bf 1}, \skip {\bf 1}, \skip {\bf 3}, \skip {\bf 3},
\skip {\bf 1},
\skip {\bf 1}, \ldots, \skip {\bf 1})\cr
           +& ( \skip {\bf 1}, \skip {\bf 1}, \skip {\bf 1}, \skip {\bf 3},
\skip {\bf 3},
\skip {\bf 1}, \ldots, \skip {\bf 1})\cr
           & \vdots \cr
           +& ( \skip {\bf 1}, \skip {\bf 1}, \skip {\bf 1}, \skip {\bf 1},
\skip {\bf 1}, \ldots,
\skip {\bf 1}, \skip {\bf 2})\}}$$
for a total of 140 charged hypermultiplets.
Again, one finds that this matter content can be obtained from the
previous one by Higgsing the $SU(4)$'s to $SU(3)$'s, yielding exactly one
extra singlet, which corresponds to the difference between the ranks of
$E_6$ and $SO(10)$.
\medskip

\subsubsection{(x)}{The mirror of the manifold with enhanced gauge group E$_7$}
The gauge group is $E_8^5{\times} F_4^4{\times} G_2^{8}{\times}
SU(2)^{17}$, and there are 61
tensors. The charged matter comes from 8 factors of $G_2^{[-3]}{\times}
SU(2)^{[-2]}$ 
plus matter that is charged under 
$SU(2)^{[-2]}$\break$ \times SU(2)^{[-2]} \times SU(2)^{[-2]} \times
SU(2)^{[-2]} \times SU(2)^{[-2]} \times SU(2)^{[-2]} \times
SU(2)^{[-2]} \times SU(2)^{[-2]} \times $\break$ SU(2)^{[-2]}$ as
follows:
$$\def\skip{\hphantom{\bf 1}}
\eqalign{\{2& ( \skip {\bf 2}, \skip {\bf 1}, \skip {\bf 1}, \skip {\bf 1},
\skip {\bf 1}, \ldots,
\skip {\bf 1}, \skip {\bf 1})\cr
           +& ( \skip {\bf 2}, \skip {\bf 2}, \skip {\bf 1}, \skip {\bf 1},
\skip {\bf 1}, \ldots,
\skip {\bf 1}, \skip {\bf 1})\cr 
           +& ( \skip {\bf 1}, \skip {\bf 2}, \skip {\bf 2}, \skip {\bf 1},
\skip {\bf 1}, \ldots,
\skip {\bf 1}, \skip {\bf 1})\cr
           +& ( \skip {\bf 1}, \skip {\bf 1}, \skip {\bf 2}, \skip {\bf 2},
\skip {\bf 1}, \ldots,
\skip {\bf 1}, \skip {\bf 1})\cr 
           & \vdots \cr
          +2& ( \skip {\bf 1}, \skip {\bf 1}, \skip {\bf 1}, \skip {\bf 1},
\skip {\bf 1}, \ldots,
\skip {\bf 1}, \skip {\bf 2})\}}$$
for a total of 104 charged hypermultiplets.
Again, one finds that this matter content can be obtained from the
previous one by Higgsing the $SU(3)$'s to $SU(2)$'s, yielding exactly one
extra singlet, which corresponds to the difference between the ranks of
$E_7$ and $E_6$.
\medskip

\subsubsection{(xi)}{The mirror of the manifold with enhanced gauge group
E$_8$}
The gauge group is $E_8^5{\times} F_4^4{\times} G_2^{8}{\times}
SU(2)^{8}$, and there are 61
tensors. The charged matter comes from 8 factors of $G_2^{[-3]}{\times}
SU(2)^{[-2]}$
for a total of 64 charged hypermultiplets.
Once again, this matter content can be obtained from the
previous one by Higgsing the $SU(2)$'s away, yielding exactly 13
extra singlets, which corresponds to the difference between the ranks of
$E_8$ and $E_7$ plus 12 extra tensor multiplets which appear in the original
spectrum when the $E_8$ is unhiggsed.
\subsection{The mirrors of models with higher values of $n$.}
It is straightforward to obtain the matter content of the mirrors of models
with higher values of $n$. We present below the matter content of the model
with $n=4$, 6 and 12. The mirror of the model with $n=12$ was also studied by
Aspinwall and Gross~
\Ref\AG{P.~S.~Aspinwall and M.~Gross, unpublished.}.
\medskip

\subsubsection{(i)}{The mirror of the model with n = 4}
The gauge group is $E_8^{9}{\times} F_4^{9}{\times} G_2^{18}{\times}
SU(2)^{18}$, and there are 107
tensors. The charged matter comes from 18 factors of $G_2^{[-3]}{\times}
SU(2)^{[-2]}$.
In addition, one of the divisors corresponding to a factor of $F_4$ has
self-intersection $-4$ (all the others have self-intersection $-5$), so that we
get additional charged matter transforming in the ${\bf 26}$ of $F_4$, 
for a total of 170 charged hypermultiplets. Note that the ${\bf 26}$ of $F_4$
contains 2 zero weight vectors, so that while there are only 6 neutral
hypermultiplets in the theory, we get two more upon going to the Coulomb
branch. This is consistent with the fact that the Hodge numbers are $(271, 7)$.
\medskip 

\subsubsection{(ii)}{The mirror of the model with n = 6}
The gauge group is $E_8^{11}{\times} F_4^{10}{\times} G_2^{21}{\times}
SU(2)^{22}$, and there are 127
tensors. The charged matter comes from 20 factors of $G_2^{[-3]}{\times}
SU(2)^{[-2]}$. In this case we also have
one factor of $SU(2)^{[-2]}{\times} G_2^{[-2]}{\times} SU(2)^{[-2]}$, 
which yields $$\{{1\over 2}({\bf 2}, {\bf 7}, {\bf 1})+{1\over 2}({\bf 2},
{\bf 1}, {\bf 1})+{1\over 2}({\bf 1}, {\bf 7}, {\bf 2})+{1\over 2}({\bf 1},
{\bf 1}, {\bf 2})+2({\bf 1}, {\bf 7}, {\bf 1})\}$$
for a total of $190$ charged hypermultiplets. Note that the ${\bf 7}$ of $G_2$
contains a zero weight vector, so that while there are only 8 neutral
hypermultiplets in the theory, we get two more upon going to the Coulomb
branch. This is consistent with the fact that the Hodge numbers are $(321, 9)$.
\medskip 

\subsubsection{(iii)}{The mirror of the model with n = 12}
The gauge group is $E_8^{17}{\times} F_4^{16}{\times} G_2^{32}{\times}
SU(2)^{32}$, and there are 193
tensors. The charged matter comes from 32 factors of $G_2^{[-3]}{\times}
SU(2)^{[-2]}$,
each of which contribute $$\{{1\over 2}({\bf 7}, {\bf 2})+{1\over 2}({\bf 1},
{\bf 2})\},$$
for a total of 256 charged hypermultiplets.
\medskip
\newpage
\section{fin}{Discussion}
In this paper, we have shown how toric geometry encodes the matter content of
F-theory compactifications on elliptic \cyt s. The Green-Schwarz
anomaly cancellation condition in six dimensions relates the matter content of
gauge theories to the geometric data~\cite{\Sadov}. We find that this relation
takes on a very simple form in terms of the toric data, making it possible to
read off the matter content from the polyhedron describing the \cyt . Thus,
in this respect, we
have almost completed the dictionary relating geometry and physics. In
particular, given any elliptic \cyt\ that has a toric description in terms of
a reflexive polyhedron, we can read
off the massless spectrum of the resulting six-dimensional theory from the
polyhedron by using the methods described in this paper and Ref.~\cite{\BG}.
All the elliptic \cyt s that are hypersurfaces in toric varieties and can be
obtained from a single weight system have already been
constructed~\cite{\AKMS}. Combined with the results of this
paper and Ref.~\cite{\BG}, this
could be interpreted as meaning that a large class of $N=1$ vacua in six
dimensions can be constructed using toric methods.

There is a point which is worth mentioning here. For every elliptic \cyt\
described by a reflexive polyhedron, the base is always a
non-singular two-dimensional toric variety. It is easy to see that all the
divisors corresponding to distinct rays in the fan of the base describe genus
zero curves. Now, in Ref.~
\Ref\KMP{S.~Katz, D.~R.~Morrison and M.~R.~Plesser, Nucl.Phys. {\bf B477}
(1996) 105, hep-th/9601108.},
it was shown that a genus $g$ curve of $A_{n-1}$ singularities in an elliptic
\cyt\ leads to an enhanced $SU(n)$ gauge symmetry with $g$ adjoint
hypermultiplets. This means that if a manifold with such a curve of
singularities can be described as a hypersurface in a toric variety, then the
divisor in the base obtained by projecting out the divisor corresponding to
the singularity cannot be seen as a ray in the base, and our methods of
determining the gauge and matter content of the theory break down. However,
this does not imply that one cannot describe vacua with adjoint
hypermultiplets using toric methods. For instance, the model with $n=2$ and
enhanced gauge group $SU(2)_c$ (in the notation of Ref.~\cite{\CF}), obtained
using toric geometry, has charged matter content consisting
of 1 {\bf 3} and 48 {\bf 2}'s, even though the genus of this curve of
singularities is zero\Footnote{We thank P.~Berglund for useful discussions on
this point.}.

There is an interesting property of the models described in \SS{4}. Consider
the Higgsing of the $E$-series
$$\def\ra{\rightarrow}
E_8\,\ra \,E_7\,\ra \,E_6\,\ra \,E_5 = SO(10)\,\ra \,E_4 = SU(5)\,\ra \,E_3 =
SU(2){\times} SU(3).$$ 
For all $n$, models with these enhanced gauge groups have mirrors whose gauge
contents form a Higgsing sequence in the opposite direction, for example, the
mirror of
the model with enhanced gauge group $E_7$ has a gauge content which can be
Higgsed to yield the mirror of the model with gauge group $E_8$ and the same
value of $n$. The actual Higgsing pattern is described in \SS{5}, for models
with $n=0$. We have found that this Higgsing pattern persists for all values
of $n$, even though the actual gauge groups are different. In all these
cases, the number of tensors in the mirror models is constant. Continuing the
Higgsing sequence further,
$$\def\ra{\rightarrow}
E_3 = SU(2)\times SU(3)\,\ra \,E_2 = SU(2)^2\,\ra \,E_1 = SU(2)\,\ra \,E_0 =
SU(1),$$
we find that the number of tensors in the mirror models increases each time,
and the mirrors are not related by Higgsing in any way.
This relation between Higgsing and unhiggsing of the models and their
mirrors is very intriguing. It would be interesting to understand the physical
meaning of this phenomenon and its relation, if any, to the duality in three 
dimensional $N=4$ theories proposed by Intriligator and Seiberg~
\REFS\IS{K. Intriligator and N. Seiberg, \plb{387} (1996) 513,
hep-th/9607207.}
\REFSCON\HOV{K. Hori, H. Ooguri and C. Vafa, hep-th/9705220.}
\refsend .
\vskip5pt
\noindent {\bf Acknowledgements}
\vskip5pt
\noindent We wish to thank A.~Avram, P.~Berglund, S.~Katz, P.~Pasanen,
E.~Silverstein and H.~Skarke for useful discussions.
This work was supported in part by the Robert Welch Foundation and NSF grant
PHY-9511632.
\newpage
\baselineskip=13pt plus 1pt minus 1pt
\immediate\closeout\referencewrite\referenceopenfalse
\line{\bf\hfil References\hfil}\bigskip\parindent=0pt\input referenc.texauxil
\bye